\def\year{2022}\relax
\documentclass[letterpaper]{article} 
\usepackage{aaai21}  
\usepackage{times}  
\usepackage{helvet} 
\usepackage{courier}  
\usepackage[hyphens]{url}  
\usepackage{graphicx} 
\usepackage{natbib}  
\usepackage{caption} 
\usepackage{subcaption}
\usepackage{todonotes}
\usepackage{amsmath}

\title{Contextualizing Online Conversational Networks}
\author{
    Thomas Magelinski and Kathleen M. Carley
    \\
}
\affiliations{
    Center for Computational Analysis of Social and Organizational Systems\\
    Carnegie Mellon University School of Computer Science\\

    5000 Forbes Avenue, Pittsburgh, Pennsylvania 15213\\
    tmagelin@cs.cmu.edu

}

\begin{document}

\maketitle

\begin{abstract}
Online social connections occur within a specific conversational context.
Prior work in network analysis of social media data attempts to contextualize data through filtering.
We propose a method of contextualizing online conversational connections automatically and illustrate this method with Twitter data.
Specifically, we detail a graph neural network model capable of representing tweets in a vector space based on their text, hashtags, URLs, and neighboring tweets.
Once tweets are represented, clusters of tweets uncover conversational contexts.
We apply our method to a dataset with 4.5 million tweets discussing the 2020 US election.
We find that even filtered data contains many different conversational contexts, with users engaging in multiple contexts.
Central users in the contextualized networks differ significantly from central users in the overall network.
This result implies that standard network analysis on social media data can be unreliable in the face of multiple conversational contexts.
We further demonstrate that dynamic analysis of conversational contexts gives a qualitative understanding of conversational flow.
\end{abstract}

\section{Introduction}
Social network analysis relies on proper contextualization of social interactions.
For offline social networks, contextualization has been traditionally done by scoping the measurement of social interactions within a physical space: a conference, an office, a school, etc.
Situating a social network within a single context both provides a clean dataset and allows for interpretable analysis.

Central members in a network of interactions within an office place can be seen as information brokers within the office.
Including information about how the workers interact outside of the office provides more information but can also muddy the analysis.
Adding out-of-office connections to the initial network is likely to affect who the central actors are, what the community structure is, and the general topology of the network.
While this denser network encodes more information, it also conflates two types of edges; workers will interact with each-other differently according to where they are.
Thus, it is more appropriate to study the \textit{contextualized} networks and the relationship between them.
This may include studying changes in centrality and community structure from one context to another.
Contextualization not only improves the specificity of the claims that can be made from the original network analysis, but also adds new information about the relationship between contexts.

\begin{figure*}
    \centering
    \includegraphics[width=0.8\textwidth]{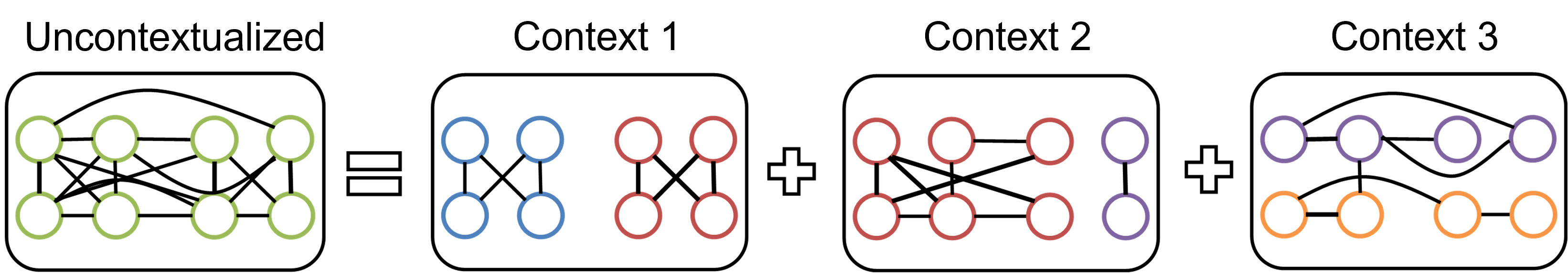}
    \caption{An illustration of different social contexts having different network structure. Before accounting for context, no community structure is apparent. After contextualization, all contexts appear to have community structure, with context 3's structure being very different from that of 1 and 2. 
    }
    \label{fig:context_cartoon}
\end{figure*}

This problem is illustrated in Figure \ref{fig:context_cartoon}. 
This simple scenario details 3 different social contexts with one having much different community structure than the others.
A standard or decontextualized analysis mixes all of the edges together and hides \textit{all} community structure.
Networks analyses are known to be sensitive to data quality, making this an important problem \cite{borgatti2006robustness}.

Methods for contextualizing social networks to date have leveraged simple property of offline networks: people can only be in one place at one time.
The result of this obvious fact is that offline social contexts occur in sequences.
For example, someone might go to work, then go to a restaurant to meet their friends, and finally return home to their family.
This then creates 3 sequential social contexts: interactions among co-workers, friends, and family.
Dynamic network analysis methods have leveraged this sequential structure can be leveraged to identify network ``states," effectively contextualizing these interactions.

Online communication is different.
Social media platforms such as Twitter are designed for users to engage in vastly different discussions simultaneously\footnote{Technically, people can only send one tweet at a time, so they are actually oscillating between conversations rather than simultaneously being engaged in them. 
The distinction for online interactions is the time-scale of state changes. 
Online discussions play-out over hours, while users switch between conversations within minutes.
This mismatch in timescales creates the ability for users to be in multiple discussions ``simultaneously"}, ruining the sequential structure of contexts.
Without this sequential structure, existing dynamic network approaches are inapplicable. 

Researchers attempt to account for this by scoping the data collection to a specific topic or event.
On Twitter, the available filters for data collection include keywords, specific users, and geographical bounding boxes.
After applying these filters, researchers can obtain reasonably contextualized datasets about a certain event or topic.
However, a related area of research, story-detection, has demonstrated that multiple events or ``stories" have separate discussions occurring even within filtered datasets.

Properly contextualizing online conversational networks is critical given their importance within the field of Social Cybersecurity \cite{carley2018social,carley2020social}.
Online conversational networks are deeply integrated into the methods for studying and understanding information operations \cite{lazer2018science,grinberg2019fake,uyheng2019interoperable}.
Thus, accurate representations of the conversational networks are necessary to understand the information space.

In this work, we propose an automated method of separating tweets into separate conversational contexts.
Just like in the offline setting, the contextualization of social interactions allows for a more precise picture of social status and group structure.
The proposed method is developed to meet the needs of our contextualization task.
Those needs are to give a vector representation of tweets which accounts for the conversational structure, works in an unsupervised setting, leverages hashtags and URLs, and works with multiple languages.
We demonstrate our approach on a real Twitter dataset consisting of 4.5 million tweets collected from November 2-8 discussing the 2020 United States Election.
We first show that the conversational contexts uncovered truly are separate discussions which should be analyzed accordingly.
Results indicate the combination of contexts leads to misleading measurements of user centrality and network structure. 
Finally, we analyze user movement between discussions, outlining the conversational flow of the entire dataset, expanding prior work which only consider contextualized networks statically and in isolation.

\section{Related Work}
Social connections must occur in the same context for social network analysis to work effectively. 
What constitutes the ``same context" depends on the study.
For example, if a study seeks to understand the spread of information in the workplace, the inclusion of connections outside the workplace may be inappropriate.
If the study instead was looking to measure epidemic spreading, all interactions are appropriate to include.
In many settings, and particularly for offline networks, this is an extremely easy requirement to meet which is easily satisfied though data collection processes such as observing connections in a specific place.

For offline networks, dynamic analysis methods have been developed to detect \textit{sequences} of network states, finding that datasets observed over longer time periods contain multiple contexts \cite{peixoto2017modelling,masuda2019detecting, magelinski2019community}.
In one example, changes can be observed from how students interact at lunch compared to in the classroom \cite{peixoto2017modelling,masuda2019detecting}.
Students interact differently in the lunch context than they can in the classroom context.
In another example, changes are observed in how Ukrainian legislatures cooperate before and after the Euromaidan revolution \cite{magelinski2019community}.
An upheaval in socio-political context disrupted friendships and rivalries between politicians.
These studies find that the community structure and central actors can be very different from context to context, and that combining contexts leads to an inaccurate representation of the network.
Accounting for contexts has also led to improvements in the modeling of processes occurring on the networks \cite{peixoto2018change}.

For online social networks, however, contextualization is not an easy task.
Two related fields have shown that social media data often contains multiple entangled contexts: topic modeling and story detection.
Topic modeling seeks to uncover a selection of different semantic contexts, or ``topics" which occur within a collection of documents \cite{blei2003latent}.
Traditional topic models such as LDA are poorly suited for the extremely short documents in Twitter data, leading to topic models specifically designed for short texts \cite{hong2010empirical,zuo2016topic,cheng2014btm}.
Alvarez-Melis and Saveski found tweets can aggregate information from their conversational context to improve topic representation \cite{alvarez2016topic}.
Other topic detection models have been developed which specifically leverage the hashtag feature of Twitter data to obtain topics \cite{wang2014hashtag,magelinski2020canadian,feng2015streamcube}.
Methods differ, but all of these works successfully demonstrate the presence of multiple semantic contexts in Twitter conversations.

Topic modeling demonstrates that entirely different things may be discussed in the same Twitter dataset, while story-detection shows that different contexts can occur even within very similar topics.
Story detection seeks to uncover ``stories" or discussions tied to specific events \cite{petrovic2010streaming,srijith2017sub,alshaabi2021storywrangler,alshaabi2021world}.
First-story detection and event-detection are very related, as they seek to identify the first tweets breaking the news of a story developing, compared to more general story-detection, which detects all the tweets in the discussion of that story \cite{walther2013geo,petrovic2010streaming,osborne2012bieber}.
In any case, detected stories are separate contexts which could otherwise be considered the same topic.
For example, story detection applied to Donald Trump's twitter timeline can distinguish within-party arguments from between-party arguments, which both belong to the topic of federal US politics \cite{dodds2021computational}.
Another example applies story detection to the Twitter discussion following the police killing of Michael Brown \cite{srijith2017sub}.
Here, fine distinctions of context are made, such as the difference in discussion of the police-lead smear campaign against Michael Brown from the discussion of the robbery that Brown committed early in the day of the shooting.
This is to say that both topic modeling and story detection develop methods of uncovering discrete conversational contexts on social media, and thereby demonstrate that these contexts exist.
These works do not, however, investigate the implications of this finding for social network analytics.

While dynamic analysis can leverage the sequential structure of human movement in offline networks, this is not possible with online networks.
The studies in topic modeling and story detection show that conversations within these conflicts can occur simultaneously, with users rapidly switching between contexts.
And while methods from topic modeling and story detection can be used to uncover conversational contexts, existing methods don't typically leverage all of the available indicators of context simultaneously: tweet text, hashtags, URLs, and the conversational graph.
Further, existing methods classify tweets with discrete labels, rather than represent them in a continuous space.
Discrete labeling is useful for network analysis, but gives no way of measuring things like distance between contexts.

Advancements in graph neural networks enable us to develop a new architecture for unsupervised Tweet representation which leverages all of the available data and places tweets in a continuous space.
Older methods of unsupervised node representation relied on random walks or ``surfs" to obtain local information which can be encoded in node vectors \cite{grover_node2vec:_2016,perozzi_deepwalk:_2014,cao_deep_2016}.
These methods do not rely on node features to obtain their representations, in contrast to the graph convolutional networks that are typically used in the semi-supervised or supervised setting \cite{kipf2016semi,hamilton2017inductive}.
Node features are necessary for tweet representation because they are used to represent the actual contents of a tweet, the tweet's text.

Methods leveraging node features have been used applied to model social media \textit{users} in a number of supervised settings, including the detection of hateful users, and the prediction of locations. \cite{ijcai2019-0881,ribeiro2018characterizing,do2018twitter}.
Perhaps the closest related model to ours is that of Nguyen et al., who used unsupervised embedding methods such as BigGraph for users, hashtags, and URLs, before combining them in a supervised retweet prediction model \cite{nguyen2020word,lerer2019pytorch}.

While models leveraging node features have been developed for social media, a mechanism for training them in an unsupervised manner was not available.
Deep Graph Infomax (DGI) filled this gap by outlining an unsupervised training procedure for feature-leveraging approaches through the  principle of mutual information \cite{velickovic2019deep}.
Similar to Stuctural Deep Network Embedding (SDNE), DGI derives an objective function in the unsupervised setting so that the architecture has something to optimize \cite{wang_structural_2016}. 

Because DGI is a methodology for training, the specific architecture for node embedding is customizable, similar to the HARP procedure \cite{chen_harp:_2017}. 
In this work, we develop a custom GCN-based architecture for representing tweets, which uses the conversational context, hashtags, and URLs, which is then trained with DGI on a real dataset.
We use the obtained tweet representations to contextualize user-to-user interactions and demonstrate the importance of contextualized network analysis.


\section{Data}
\subsection{Data Collection}
The data collection strategy for this study is intended to match the typical procedures used in the field of Social-Cybersecurity, which relies heavily on network analysis of social media discussions \cite{uyheng2019interoperable}.
Thus, the data was captured using a keyword-based stream of Twitter's API from November 2 2020 to November 8 2020. This allowed for the capture of data one day before election night, which was November 3 2020, and one day after major news outlets declared Joe Biden the winner on November 7 2020.
The keywords\footnote{\#election2020, \#presidentialelection, \#JoeBiden, \#Biden, \#BidenHarris2020, \#MAGA, \#KAG, \#democrats, \#republicans, \#VoteByMai, \#USPS, \#SaveTheUSPS, \#voterfraud, \#reopen, \#reopenamerica, \#BLM, \#BlackLivesMatter, \#QAnon, \#WWG1WGA, \#IranSanctions, ``natural born," and relevant politician's handles.} were selected in order to maximize conversation around the election.
This includes general hashtags, campaign hashtags, and mentions of prominent election figures such as Trump, Pence, Biden, and Harris.
It also includes hashtags relating to anticipated election-related issues, such as the Black Lives Matter movement, US Sanctions on Iran, issues with voting-by-mail, and claims of voter fraud.
The collection resulted in 4.5M tweets, 75k hashtags, and 47k URLs. 
The dataset approximately contains 2M retweets, 1.3M quotes, and 1.3M replies.
Hashtags were used in tweets 886k times, whereas URLs were used in tweets 75k times.

\subsection{Data Cleaning}
Tweet text was cleaned by first removing all URLs, hashtags, and mentions.
Next, punctuation was removed.
Finally, text was tokenized in preparation for the text embedding discussed in the Methodology section.

The procedure for URL normalization was as follows.
First, text before the domain name was removed.
Next, URL parameters were removed for links with domains other than ``facebook", ``google", and ``youtube."
These parameters commonly store information about the user who shared the link, among other things. 
The presence of these parameters prevents direct matching between URLs.
For ``facebook", ``google", and ``youtube," however, these parameters are used to point to the actual destination, so cannot be removed.
``Amp" links were converted to non-amp links.
Lastly, youtube.com, and yout.be links were all converted to the yout.be format.

All links to twitter.com were not considered to be typical URLs, as they are either links to media or quotes of other tweets.
Links to media were not included, while the metadata from quote-links was used to add the appropriate quote-edges in the tweet-tweet network discussed below. 
Hashtags were lower-cased, as case does not affect their functionality.

\subsection{Heterogeneous Network Construction}
We construct our heterogeneous Twitter network with three node types (tweet, hashtag, and URL) and three edge types (tweet-URL, tweet-hashtag, and tweet-tweet).
When a URL or hashtag is used in a tweet, an \textit{undirected} edge is drawn between them.
The selection of an undirected edge allows for URLs (and hashtags) to aggregate information from all the tweets they appear in, while allowing tweets to aggregate information from the URLs (and hashtags) they contain.

The third relationship, tweet-tweet, occurs through replies or quotes.
While these are slightly different operations, they both create the effect of continuing the conversation with a new tweet connected to the original.
Future work may investigate leveraging slight differences between replies and quotes.
Edges between tweets can be modeled as directed \textit{or} undirected, as a setting within the model.
A directed edge allows the reply or quote's representation to be affected by the original tweet's representation while keeping the original tweet's representation isolated.
This is an intuitive modeling approach; however, modeling this relation with an \textit{undirected} edge allows for base tweets to obtain some context, which can push similar but disconnected conversations closer together.
The undirected approach was selected because it showed better results in our experiments.

Retweets are simply copies of tweets, so they will provide no additional information from a tweet-representation point of view.
Worse, they are such a large fraction of the dataset that they could have adverse effects on the training process. 
Instead, we give retweets the same representation as their original tweet.
Thus, retweets will always be considered in the same context as the original tweet, where the user-to-user implications of retweets are studied.

\section{Methodology}
\subsection{Tweet Text Embedding}
Graph convolutional networks require some form of node-features.
We derive features for tweets using the tweet text.
To limit the scope of analysis to our proposed architecture and to enable the use of multi-language text embedding, we used the pre-trained\footnote{https://fasttext.cc/docs/en/aligned-vectors.html} and language-aligned vectors trained using fastText on the Wikipedia corpus \cite{joulin2018loss,bojanowski2017enriching}.
The use of language-aligned vectors allows us to place similar tweets in the same discussion, even if they are tweeting in different languages.

We rely on the Twitter language detection output for the classification of tweet language.
Many tweets, however, do not have an available language label.
This often occurs when tweets do not have text, but instead only have URLs, emojis, images, and sometimes hashtags.
In our case, 15.6\% of tweets in the dataset do not have an available label, and therefore cannot be embedded with this approach. 

For each tweet with a label, we perform a normalized tf-idf weighting of the fastText word vectors to obtain a 300-dimensional tweet-text embedding.
The tf-idf weights were calculated within-language to prevent language-based abnormalities.
We use this procedure to embed tweets in Arabic, English, French, German, Hebrew, Italian, Portuguese, Romanian, Russian, Spanish, and Turkish, covering over 95\% of the reachable tweets.

Finally, we use feature propagation to obtain a feature vector for the remaining tweets \cite{rossi2021unreasonable}.
Feature propagation holds known feature vectors fixed while iterative updating unknown feature vectors. 
In each iteration, each node with an unknown feature vector updates its vector by taking the average features of its neighbors.
Nodes with unknown features which have not been reached by the propagation are not counted in the update step.
After few iterations, all features converge.
Rossi et al. demonstrate that this approach yields good results in downstream tasks such as node classification even in the face of extreme missing data, when 99\% of nodes are featureless.
The task of filling in features for 20\% of nodes is much less formidable.
Feature vectors converged within 40 iterations on our dataset.

\subsection{Deep Tweet Infomax}
\begin{figure}
    \centering
    \includegraphics[width=0.65\columnwidth]{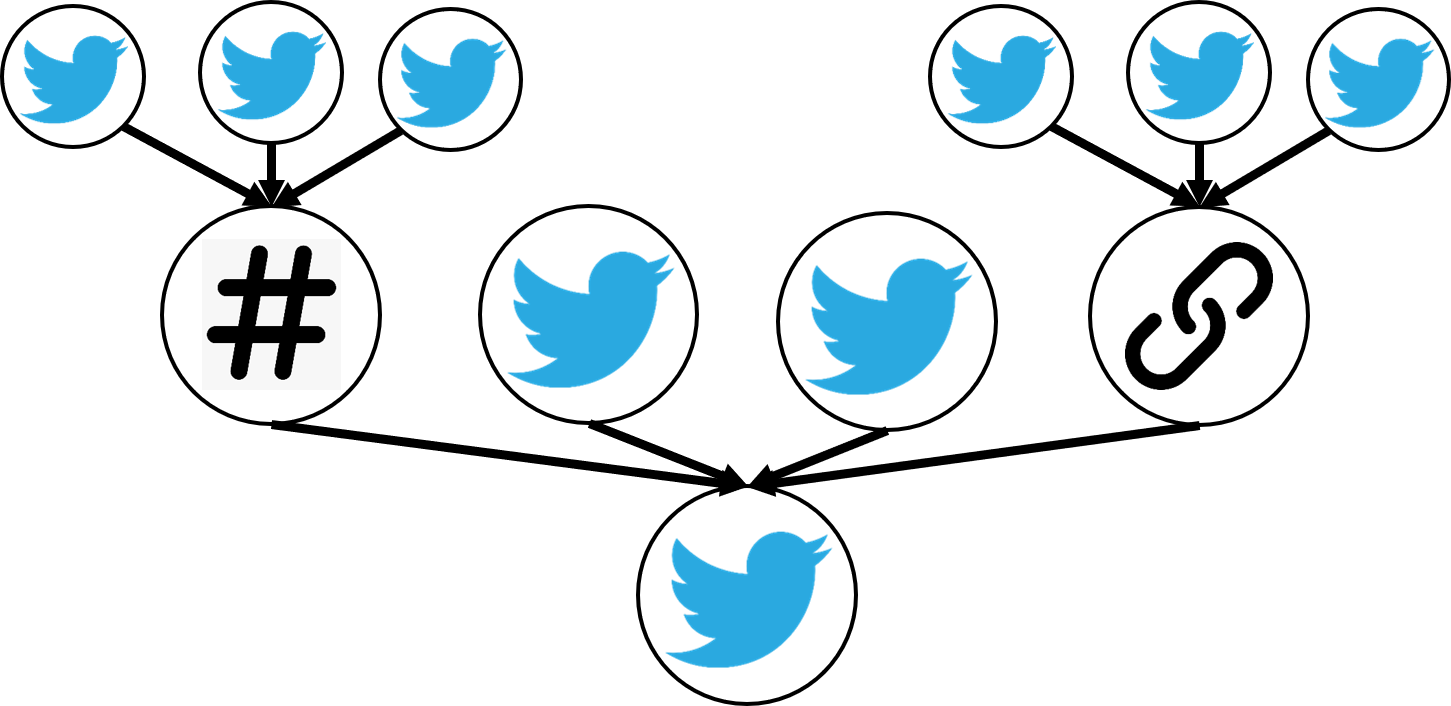}
    \caption{The flow of information in one layer of DTI. The tweet being represented is shown at the bottom. It contains a hashtag, a URL, and is connected to two tweets via reply or mention. First, the Hashtag and URL obtain a representation by aggregating from tweets that the Hashtag and URL appear in, respectively. Then, the target tweet aggregates information from all its neighbors, the tweets, the hashtag, and the URL, to obtain its representation.
    }
    \label{fig:dti}
\end{figure}

We propose an unsupervised approach for tweet representation.
The flow of information in 1 step of the graph neural network architecture can be seen in Figure \ref{fig:dti}.
A tweet aggregates information from tweets that it is connected to (replies, or quotes in either direction), hashtags, and URLS.
Hashtags and URLs, obtain information by aggregating from the tweets that they are used in.
This approach allows for all tweets using the same hashtags and URLs to pass information to one another in a memory-efficient manner, while obtaining hashtag and URL representations simultaneously.
This model is trained using Deep Graph Infomax, leading to the informal name of our approach of Deep Tweet Infomax (DTI).
The architecture will now be described in detail.

Let $t$, $u$, and $h$ represent nodes of the type tweet, URL, and hashtag, respectively.
They will be indexed using subscripts, e.g., $t_i$ corresponds to the $i^\text{th}$ tweet.
Feature vectors are represented with the letter $x$, using subscripts to indicate the corresponding node and superscripts to indicate the layer.
For example, $x_{t_i}^0$, represents the $0^\text{th}$-layer vector (otherwise known as the feature vector) for the $i^\text{th}$ tweet.
We will make use of a neighborhood function $\mathcal{N}$, which takes in a node and returns the set of its neighbors.
Subscripts of the neighborhood function allow for the return of only a specific type of neighbor.
For example, $\mathcal{N}_u(t_i)$ returns all of the URLs connected to the $i^\text{th}$ tweet.

Hashtags and URLs themselves do not have an initial feature representation.
Instead, they aggregate the feature representation from all of the tweets that they are used in.
As tweet-representation improves, so does hashtag and URL representation.
Thus, the first step of the representation process is for hashtags and URLs to aggregate information from the tweets that they appear in, as seen in Equations \ref{eq:hash_agg} and \ref{eq:url_agg}, where AGG is a learnable aggregation function, and $\sigma$ is an activation function.

\begin{align}
    x_{h_i}^0 &= \sigma(\text{AGG}(\{x_{t_i}^0, \forall t_i \in \mathcal{N}_t(h_i) \})) 
    \label{eq:hash_agg}
    \\ 
    x_{u_i}^0 &= \sigma(\text{AGG}(\{x_{t_i}^0, \forall t_i \in \mathcal{N}_t(u_i)\}))
    \label{eq:url_agg}
\end{align}

Now that all nodes have feature vectors, tweets aggregate from their heterogeneous neighborhoods.
Separate aggregation functions are learned for the tweets, hashtags, and URLs that a tweet is connected to, which are then averaged, and an activation function is applied, as seen in Equation \ref{eq:tweet_agg}.

\begin{align}
\label{eq:tweet_agg}
    \begin{split}
        x_{t_i}^1 &= \sigma( \frac{1}{3} ( \text{AGG}(\{x_{h_i}^0, \forall h_i \in \mathcal{N}_h(t_i) \})
        \\
        &+ \text{AGG}(\{x_{u_i}^0, \forall u_i \in \mathcal{N}_u(t_i) \})
        \\
        &+ \text{AGG}(\{x_{t_i}^0, \forall t_i \in \mathcal{N}_t(t_i) \cup \{t_i\} \}) ) )
    \end{split}
\end{align}

The process thus far defines the network over which features are passed, and the order in which to pass them.
The selection of the aggregation function, $\text{AGG}$, is the main topic of debate within graph neural network research.
In future work, $\text{AGG}$, can be expected to be substituted for the new state-of-the-art aggregation schemes.
We employ the GraphSAGE aggregation, which is the initial aggregation scheme applied in the Deep Graph Infomax work \cite{hamilton2017inductive}.
This aggregation scheme is detailed for the tweet-to-tweet relationship in Equation \ref{eq:graphsage}, where \textbf{W} are trainable weight matrices, and \textbf{b} is a trainable bias vector.
Note that the first term is only present when a feature vector from the previous layer is available.
So for the initial aggregation steps of hashtags and URLs, this term is not present.

\begin{align}
    x_{t_i}' &= \mathbf{W}_1 x_{t_i}^0 + \frac{1}{|\mathcal{N}(t_i)|} \sum_{t_j\in \mathcal{N}(t_i)}\mathbf{W}_2 x_{t_j}^0 + \mathbf{b}
    \label{eq:graphsage}
\end{align}

GraphSAGE is a relatively simplistic aggregation scheme, where all neighbors are treated equally.
More recent aggregation schemes add attention, which allows for a weighted average of neighbors.
We also trained DTI using the attention function detailed by Brody, Alon, and Yahav, which improved on the original graph-attention network from Veli{\v{c}}kovi{\'c} et al \cite{brody2021attentive,velivckovic2017graph}.
However, GraphSage out-performed aggregation with attention in our experiments, so results are given for the GraphSAGE implementation.

Finally, we must select a nonlinear activation function.
Again following the original Deep Graph Infomax work, we use the PReLU, activation function \cite{he2015delving}.  


The process up to here details a single-layer of the architecture.
Tweets will only obtain information from 1-hop away, and hashtags and URLs will only receive information from the initial feature vectors.
Stacking these layers enables further information spread, and thus representations or tweets, URLs, and hashtags.
Here, we stack two of these layers.
Classically, a depth of 2 is very shallow.
However, in our case, tweet networks themselves are shallow. 
The vast majority of Twitter replies are replies to a base-tweet, rather than replies to replies.
Our choice of encoding tweet relationships with undirected edges also informs this depth.

To train this architecture, we use Deep Graph Infomax (DGI), an approach for learning unsupervised node representations by maximizing mutual information between patch representations and corresponding high-level summaries of graphs \cite{velickovic2019deep}.
We note that a version of DGI has been developed specifically for heterogeneous networks \cite{ren2019heterogeneous}.
Because of our focus on tweet representations and the lack of features available for URLs and hashtags, we proceed with the original formulation of DGI.

The DGI training process involves four steps.
First, a normal forward pass on the data is performed, giving tweet representations, $\mathbf{x_t}$.
Next, a readout function is applied to give a graph-level summary vector, $\mathbf{s}$.
Velickovic et al. applied a sigmoid function to a simple averaging of the node vectors but suggest that  more sophisticated methods such as the Set2Vec method developed by Vinyals et al. could perform better on larger graphs \cite{vinyals2015order}.
In our experiments this does seem to be the case, so we apply Set2Vec using 5 processing steps: $\mathbf{s} = \sigma(\text{Set2Vec}(\{x_{t_i} \forall {t_i}\}))$, where $\sigma$ is the sigmoid function.
Third, a forward pass is performed on corrupted data, giving corrupted tweet representations, $\mathbf{\Tilde{x}}_t$.
We use the same corruption function as the original work, a shuffling of the tweet features while keep edges intact.
Finally, to classify tweets as corrupted or not a scoring function is given as $d_{t_i} = \sigma(x_{t_i}^T\mathbf{W}s)$, where \textbf{W} is a trainable scoring matrix and $\sigma$ is the sigmoid function, providing a score between 0 and 1.
Binary cross entropy loss was used on the score, $d$, and the label (corrupted or not) to train the model. 

The model was implemented using the PyG library \cite{fey2019fast}.
All hidden and output layer dimensions were set to 128.
The model was trained using mini-batches of 24000 tweets, corresponding to 100 mini-batches per epoch.
PyG's ``NeighborLoader" was used to handle the neighborhood sampling within minibatches, where 20 neighbors of each edge-type were sampled for 3 iterations.
The ADAM optimizer was used during gradient descent with an initial learning rate of $\alpha=0.001$ for 25 epochs with early stopping \cite{kingma2014adam}.
The model trained in about 12 hours on an Intel E5-2687W v3 CPU.

\subsection{Clustering}
Once tweets are represented in a continuous space through DTI, they can be clustered with a variety of clustering algorithms.
Tweet clusters, then, are the discrete contexts that conversational networks can be studied within.
Given the size of Twitter datasets and the unknown number of desired clusters, density-based clusters like DBSCAN are a reasonable choice \cite{ester1996density}.

Twitter datasets tend to be large, and although density-based clustering algorithms are highly optimized, it is extremely costly to run them many times to select appropriate parameters as was done in Schubert et al
To minimize the need for parameter tuning when clustering, we use the hierarchical version of DBSCAN, HDBSCAN, which requires less parameters \cite{mcinnes2017hdbscan,mcinnes2017accelerated}.
We perform clustering with 1 minimum samples in order to cluster the maximum number of tweets, and a minimum cluster size of 100.

\section{Results}
\subsection{Validity of Embeddings}
We validate our tweet representations in two steps.
First, we use a simplistic data annotation scheme and see how 5 categories of tweets fall within the embedded space.
We find that the clusters in the tweet-embedding space are well-correlated with the annotated groups.
Second, we detail some of the URL and hashtag's nearest-neighbors to demonstrate that intermediate steps within the model are working.

The 100 URLs receiving the most cumulative retweets in our dataset were hand-labeled with the story that they pertain to.
This resulted in 35 story labels.
The most popular category was spam as spammers often tweet the same URL many times.
The following 5 most popular stories labeled 53 of the remaining 80 URLs, returns diminished significantly when including more than these 5 stories.
These stories include ``Election Updates," ``Pro-Trump Conspiracy," ``Vote Info," ``Vote Biden," and ``Vote Trump."
To clarify, URLs with the  ``Election Updates" were those giving updates to the election; ``Vote Biden/Trump" URLs gave reasoning to vote for Biden/Trump, ``Vote Info" URLs gave information on how to vote, and ``Pro-Trump Conspiracy" URLs include several spurious conspiracy theories which lean in Trump's favor.
Most of these conspiracies give false or misleading information in support of the false premise that the Democrats stole the 2020 election from Trump.

\begin{figure*}
    \centering
    \begin{subfigure}[b]{0.4\textwidth}
         \centering
         \includegraphics[height=2.25in]{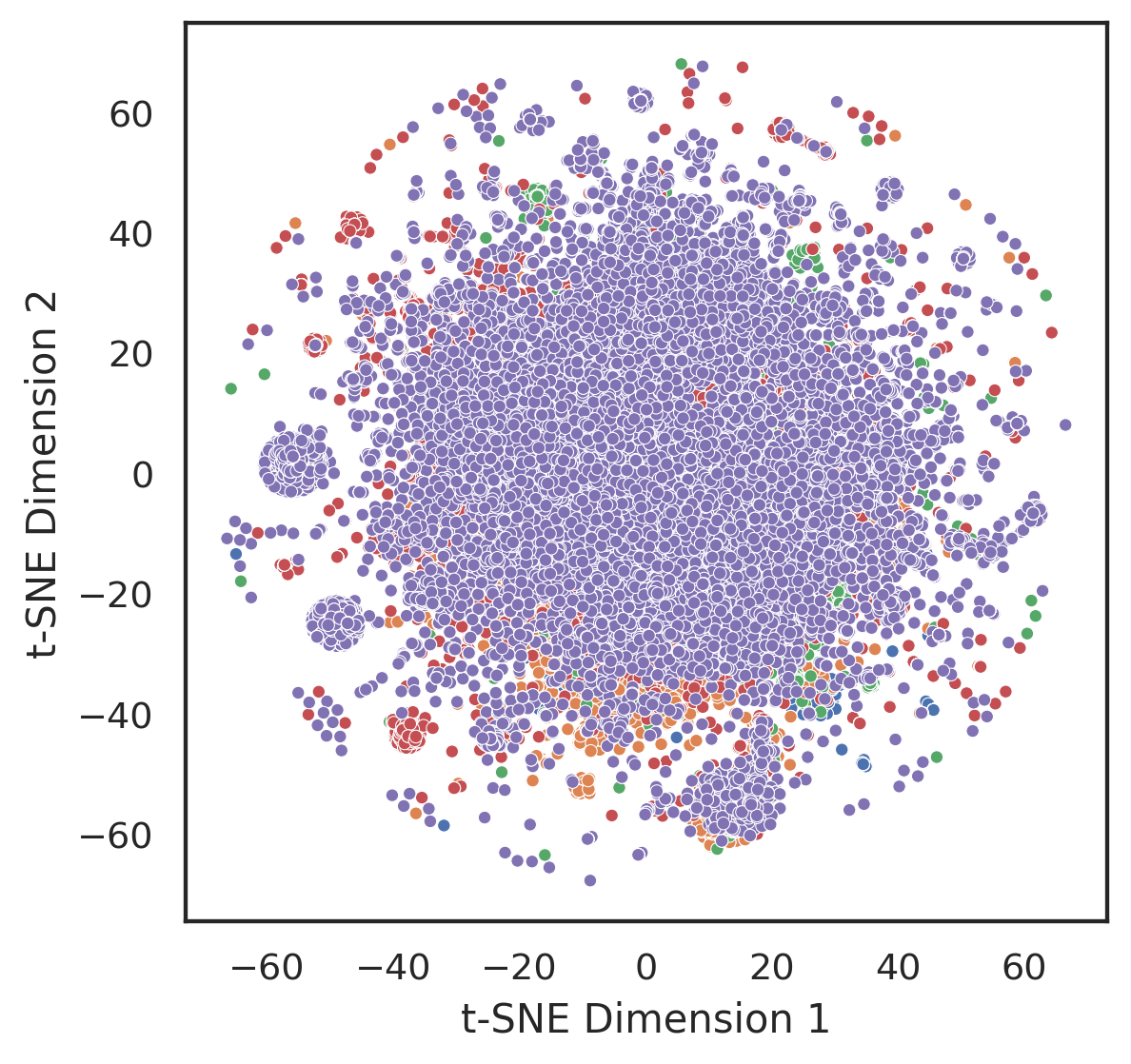}
         \caption{Text Embedding}
         \label{fig:text_tnse}
    \end{subfigure}
     \hfill
    \begin{subfigure}[b]{0.55\textwidth}
         \centering
         \includegraphics[height=2.25in]{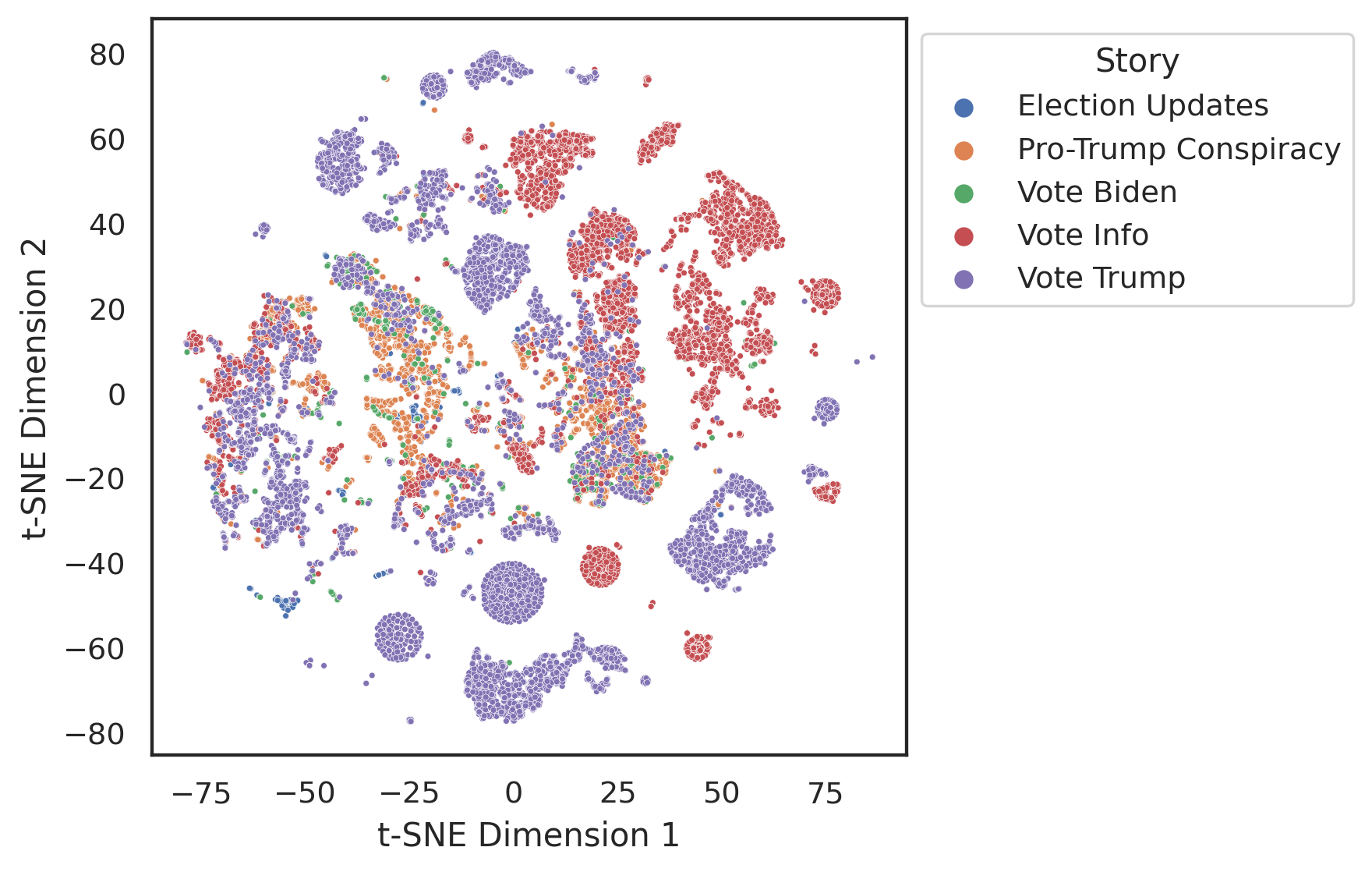}
         \caption{DTI Embedding}
         \label{fig:dti_tnse}
    \end{subfigure}
    \caption{2-dimensional t-SNE representation of tweets before and after DTI training. These are the tweets within 2-hops of a hand-annotated URL. The tweets are colored with the annotation of their associated URL.}
    \label{fig:tweet_story_tsne}
\end{figure*}

The labeled URLs were then used to annotate tweets through a 2-step label propagation.
This process assumes that tweets using a URL are part of the discussion that URL refers to, and that tweets directly connected to a tweet with URL are also part of that discussion.
Further propagation is possible, but the assumption that tweets further and further from the initial URL should still have the same label becomes harder to justify.
Tweets that have conflicting labels were not included, though these made up less than 1\% of the tweets.
Each labeled tweet was projected into a 2-dimensional space using t-SNE on their initial text embeddings and their final embeddings in Figures \ref{fig:text_tnse} and \ref{fig:dti_tnse}, respectively, where tweets are colored by their label \cite{van2008visualizing}.


The text embedding in Figure \ref{fig:text_tnse}, is similar to Sia et al.'s approach to topic modeling and is used as a baseline \cite{sia-etal-2020-tired}.
We observe that the text embedding is unable to recover the conversational contexts we set out to find.
This is likely due to the facts that these contexts have similar word distribution, and that a text-only approach cannot leverage replies, hashtags, or URLs. 
Next, we observe in Figure \ref{fig:dti_tnse} that there are a number of well-formed tweet clusters which  correspond to different conversational contexts.
We observe that some of these clusters form tight balls, almost perfect circles in the embedded space.
Investigation into these regions finds that this occurs when many tweets reply or quote a popular tweet.
The original tweet anchors the conversation, while the additional information in replies or quotes move these secondary quotes in different directions within the embedded space, but not far from their neighboring tweet.
We also see that not all of the clusters are so simple, pointing to more interesting contextual structure.

Importantly, the majority of tweet clusters have homogeneous labels.
We see many clusters with 100\% label agreement labeled with ``Vote Trump," and ``Vote Info."
There are clusters with noticeably higher than average density of ``Election Updates" and ``Vote Biden" tweets, but they are surrounded by ``Pro-Trump Conspiracy" tweets, and occasionally ``Vote Trump" tweets.
This follows from the observation that many tweets about the election or statements in support of Biden were replied or quoted with lies about voter fraud and the Democrat's efforts to steal the election.

To dive deeper into the specifics of the embeddings, node's nearest neighbors were analyzed.
Specifically, the top-5 closest pairs of hashtags and URLs are displayed in Tables \ref{tab:hashtag_neighbors} and \ref{tab:url_neighbors}, respectively.
The nearest-neighbor analysis only considers the top-500 nodes in terms of their cumulative retweets in the dataset.
Distance is calculated as euclidean distance in the 128-dimensional space.

For hashtags, we observe a tight triangle of hashtags all used in a Japanese discussion about how Trump won and how Japan has prepared for the moment.
Next is a pair of hashtags relating to an alleged massacre in Obigbo, Nigeria. One labeling the massacre and another blaming the Nigerian Governor.
Finally, the acronyms for battleground states Pennsylvania and North Carolina appear as a pair.

\begin{table}[htb]
    \centering
    \begin{tabular}{|c|c|c|}
        \hline
        \textbf{Hashtag 1} & \textbf{Hashtag 2} & \textbf{Distance} \\
        \hline
        \#returnoftheusa & \#japanisready* & 0.006 \\
        \hline
        \#returnoftheusa & \#trumpwins* & 0.013 \\
        \hline
        \#trumpwins* & \#japanisread* & 0.018 \\
        \hline
        \#obigbomassacre & \#govwikethemurderer & 0.023 \\
        \hline
        \#pa & \#nc & 0.026 \\
        \hline
    \end{tabular}
    \caption{Pairs of hashtags that are closest in the embedded space. Only the top 500 hashtags are considered, as they are the most important for tweet representation and are the cleanest. A star indicates translation from Japanese.
    }
    \label{tab:hashtag_neighbors}
\end{table}

\begin{table*}[htb]
    \centering
    \begin{tabular}{|p{0.4\linewidth}|p{0.4\linewidth}|p{0.1\linewidth}|}
        \hline
        \textbf{URL 1} & \textbf{URL 2} & \textbf{Distance} \\
        \hline
        \url{aje.io/3p45z} & \url{aje.io/rlmfd} & 0.050 \\
        \hline
        \url{theredelephants.com/there-is-undeniable-mathematical-evidence-the-election-is-being-stolen} & \url{noqreport.com/2020/11/04/hammer-and-scorecard-lt-gen-mcinerney-explains-the-election-hack-by-democrats} & 0.072 \\
        \hline
        \url{nypost.com/2020/08/29/political-insider-explains-voter-fraud-with-mail-in-ballots} & \url{technocracy.news/the-hammer-and-scorecard-weapons-of-mass-vote-manipulation} & 0.076 \\
        \hline
        \url{theredelephants.com/there-is-undeniable-mathematical-evidence-the-election-is-being-stolen} & \url{youtu.be/ivT2z5UgHxo} & 0.077 \\
        \hline
        \url{zerohedge.com/political/michigan-usps-whistleblower-claims-late-ballots-backdated} & \url{nytimes.com/2020/11/07/us/politics/theres-no-evidence-to-support-claims-that-election-observers-were-blocked-from-counting-rooms.html} & 0.077 \\
        \hline
    \end{tabular}
    \caption{Pairs of URLs that are closest in the embedded space. Only the top 500 URLs are considered, as they are the most important for tweet representation, and are the cleanest.}
    \label{tab:url_neighbors}
\end{table*}

For URL neighbors, we first see a pair of Aljazeera links.
The first is the live election results board, while the second is the live discussion of election updates.
The next three pairs of links are pushing the false narrative of election fraud and vote manipulation. The YouTube link points to a video of Steven Crowder discussing voter fraud in Michigan.
Lastly, we see two URLs with differing objectives, one making claims of voter fraud, and another fact-checking a claim of vote interference.
This pair demonstrates that when competing claims are made in the same conversations, opposing URLs or hashtags can be close in the embedded space.

The neighbors in both URL and hashtag space seem to be well-matched pairs.
This gives validity to the methods ability to represent information using the network's connections.
It also gives validity to the tweet embeddings, because they directly rely on the representations of URL and hashtags.

\begin{table*}[htb]
    \centering
    \begin{tabular}{|p{0.46\linewidth}|p{0.46\linewidth}|}
        \hline
        \textbf{Tweet 1} & \textbf{Tweet 2} \\
        \hline
        breaking: fbi reports military mail-in ballots were discarded in pennsylvania. all cast for @realdonaldtrump voter fraud is an epidemic in the democrat party! it is up to americans to restore integrity in our election process. & According to a statement released by the fbi, several military ballots were found dumped in a ditch in Pennsylvania. All of them were votes for President Trump. This is the real election obstruction. Not reported in Japan at all! (ja)  \\
        \hline
        ``President Trump, as a friend, I sincerely pray that Mrs. Melania will be completely recovered as soon as possible." @realDonaldTrump & Biden slashed Trump to become president 46th place, after the people all over the world are excited -- , we will start our first program by sitting with US political professor and then dissected to hear that the United States and What will the world system be like in the Biden era?  please follow.  @JoeBiden wins; he will be the 46th POTUS. (th) \\
        \hline
        (1) ``I can't breathe" is the sentence George Floyd calls for help. After being hit by a white policeman using his knees to press his neck for a long time until he finally dies And that became the trigger to awaken the trend \#BlackLivesMatter , but in Thailand it has happened like that (continued). (th) & Here are 11 overseas video works that may help you to understand the essential background of the \#BlackLivesMatter movement, which has become a huge trend mainly in the United States. We have carefully selected works that can be viewed on the web, such as Netflix and Amazon Prime (unlimited viewing / rental), so please take this opportunity to check them out. (ja) \\
        \hline
    \end{tabular}
    \caption{Pairs of different-language tweets that are closest in the embedded space. Only the top 1000 tweets which had did not have ``undefined" language were considered. Translated with Google Translate from languages codes appended to the quote. 
    }
    \label{tab:tweet_neighbors}
\end{table*}

Lastly, the closest pairs of different language tweets in the embedding space are given in Table \ref{tab:tweet_neighbors}.
Off-language pairs were chosen to highlight the method's ability to work in the multi-lingual setting, while the overall neighbors are omitted given space restrictions.
The first and second pair of tweets are extremely similar tweets that differed in language.
The second pair has different semantic meaning, but from a bag-of-words perspective we can understand that these two should be fairly close in the embedded space.

\subsection{User Overlap in Tweet Contexts}
Because of the size of the dataset, clustering all of the tweets simultaneously is very costly.
Instead, we cluster tweets on a daily basis.
Conversations evolve so quickly on social media, that it is reasonable to assume that most clusters will be with tweets occurring at roughly the same time.

\begin{figure}
    \centering
    \includegraphics[width=0.8\columnwidth]{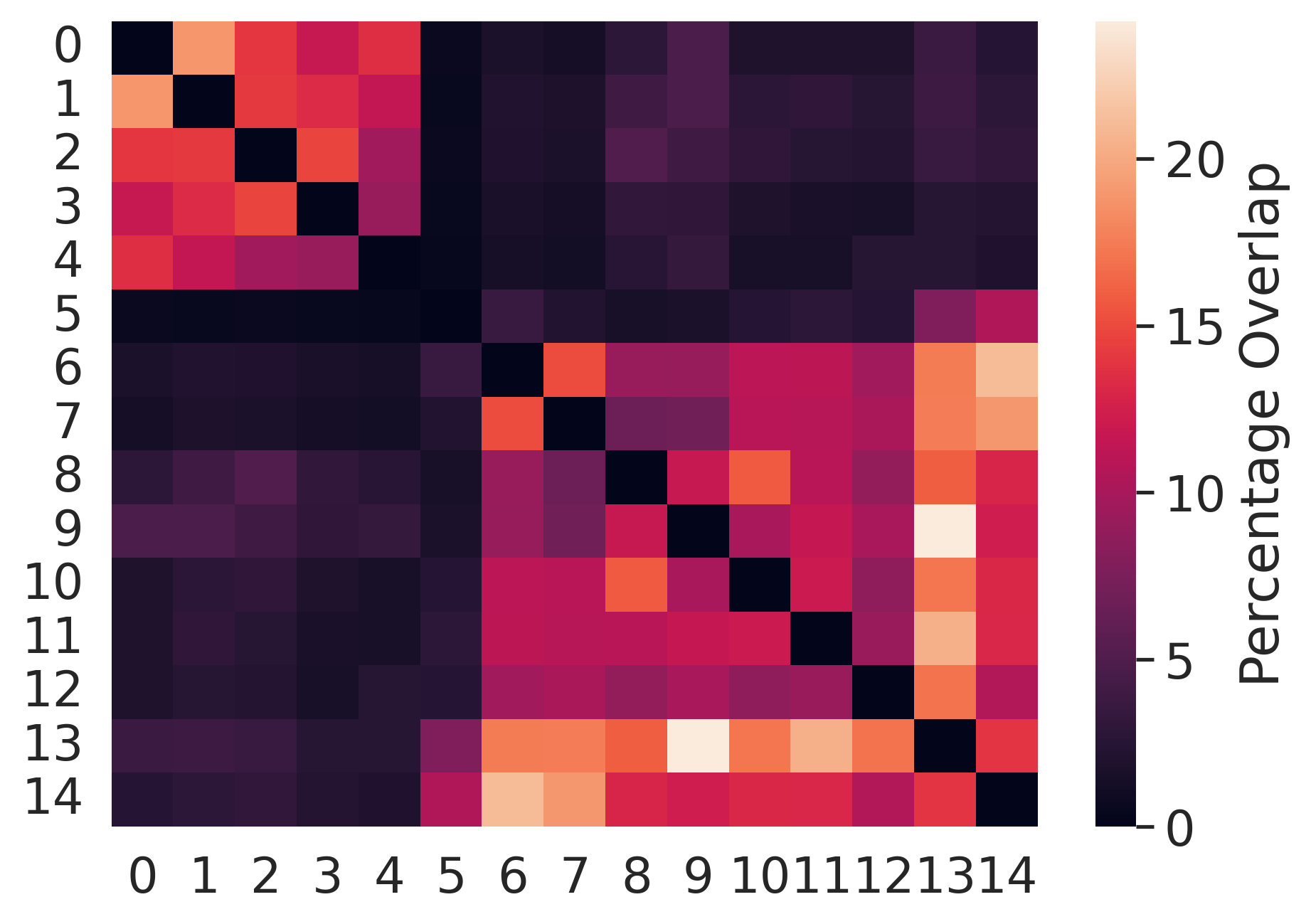}
    \caption{Overlap of active users in the 15 largest conversational contexts. The diagonal is set to 0\% for readability.}
    \label{fig:conv_overlap}
\end{figure}

We consider users to be active within a conversational context if their tweet is in the context, or they retweet one that is.
For the largest 15 conversational contexts, we calculate the pairwise percentage of overlap in membership and plot the results in Figure \ref{fig:conv_overlap}.
We see that there are two context clusters which have little no member overlap.
Within each cluster there is higher overlap, but still only around 15-20\%.

This finding has important ramifications for conversational network analysis.
The presence of non-overlapping contexts highlight that \textit{global} properties of conversational networks are being affected by context. 
Placing users from the first cluster of contexts in the same network as those in the second is a misleading representation of the data.
It is possible that users from these different contexts may even be placed in the same component of a decontextualized network.
As the number of active users increases, it becomes more likely that the two contexts will be merged into a single component under decontextualized analysis. 

More importantly, there are contexts will low but not negligible overlap, around 15\%.
This means that the \textit{local} properties of the de-contextualized network are affected.
With 15\% overlap, we can expect that about 15\% of users will have connections from users in both contexts with no way of distinguishing them.
This has negative effects on every aspect of network analysis.
Path-based centralities, for example, will be calculated on paths that could not occur in the data because they span contexts.
The impact of this is further studied in the following section.

\subsection{The Importance of Contextualization}
Here, we demonstrate the importance of contextualizing our conversational networks.
For the two largest clusters, conversational networks were constructed.
These are user-user networks, where connections are drawn if a user is mentioned, retweeted, replied-to, or quoted.
The overlap between these two contexts is 13.9\%.
Next, the top-5 most central users are found according to weighted PageRank centrality \cite{page1999pagerank}.
Finally, the contextualized networks are combined, and centralities are recalculated.
The percentile of centrality in the combined context for the central members is given in Table \ref{tab:context_cent}, higher being more central.
We see that the importance of top-ranked nodes in contextualized networks is diluted when contexts are not considered.
Some nodes remain relatively central, within the $80$th percentile of nodes, while others fall into the bottom quartile of centrality.
Further, we apply Kendall-Tau rank correlation to measure the correlation of centrality measures overall, not just for top influencers.
We find no significant correlation between the importance of nodes in the contextualized case versus their ranking in the decontextualized case.
Based on these results, we see that ignorance to conversational context results in a corrupted network analysis.

\begin{table}[htb]
    \centering
    \begin{tabular}{|c|c|}
        \hline
        \textbf{True Rank} & \textbf{Corrupted Percentile} \\
        \hline
        $1_1$ & 62.7\%\\ 
        \hline
        $2_1$ & 88.3\%\\ 
        \hline
        $3_1$ & 26.1\% \\ 
        \hline
        $4_1$ & 33.9\% \\ 
        \hline
        $5_1$ & 48.3\%\\ 
        \hline
        $1_2$ & 62.8\% \\ 
        \hline
        $2_2$ & 88.3\% \\ 
        \hline
        $3_2$ & 33.9\% \\ 
        \hline
        $4_2$ & 23.6\% \\ 
        \hline
        $5_2$ & 96.2\% \\ 
        \hline
    \end{tabular}
    \caption{Pagerank centrality ranking for the top-5 most central users in two contexts. True rank indicates their contextualize rank, with the subscript indicating the corresponding context they belong to. Corrupted percentile indicates their rank in the uncontextualized setting.}
    \label{tab:context_cent}
\end{table}

\subsection{Contextual Dynamics}
\begin{figure}[htb]
    \centering
    \includegraphics[width=0.8\columnwidth]{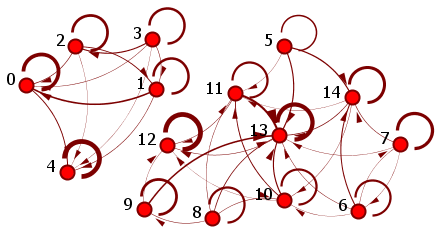}
    \caption{The user transition network between conversational contexts. Edge weight indicates probability of transition from one context to another. Edges with probability below 5\% have been trimmed for visibility.}
    \label{fig:context_dynamics}
\end{figure}

In addition to giving more accurate conversational networks, contextualization allows us to uncover new aspects of our data.
Previously, ``topic-groups" or communities formed around a topic were studied in statically and in isolation, however we can now use the patterns of user transitions between contexts to map out the flow of conversation \cite{himelboim2017classifying,smith2014mapping}.
First, users are situated in contexts when they tweet, or retweet within that context.
Then, \textit{transitions} between contexts are recorded. 
This creates a transition network from context to context.
normalizing the outward flow of this network to 1 yields a Markov transition matrix between contexts.
This transition network between the 15 largest contexts is visualized in Figure \ref{fig:context_dynamics}, where probabilities below 5\% are trimmed.

In Figure \ref{fig:context_dynamics}, we observe rich conversational dynamics.
We see two separate dynamic processes, corresponding to the different context clusters previously observed.
Two phenomena observed in context transition dynamics are of interest.
First, some contexts, such as context 6 and context 7 contain a cycle, meaning that users transition back and forth between them.
This is indicative of one of the motivating patterns observed in social media, where users can switch back and forth between contexts.
Second, we see a number of ``sinks" or conversational contexts that pull in users from 
several contexts.
The characteristic property of sinks is many heavily weighted ``in" links, corresponding to a high probability of users joining the conversational context.
Here we observe states 0, 4, 12, and 13 to be sinks.
To some extent, the presence of sinks are a function of the temporal ordering of contexts enforced due to computational limitations.
Contexts occurring later in the timeline have a better chance of being a sink than those occurring earlier.
Despite this, the observed dynamics are much more complex than a simple sequence.

\section{Discussion}
Tweet representation through Deep Tweet Infomax appears to be well-validated.
This validation stems from a 3-part analysis, where tweets, URLs, and hashtags were all tested.
URLs and hashtags were observed to have meaningful nearest neighbors in the embedded space, despite being an intermediate step in the tweet embedding process.
Tweets, annotated with nearby URLs (which were themselves hand-labeled), form clusters of the same labels within the embedded space.
With this validation in mind, we move to discussion of the network results.

The observed conversational contexts separate well into 2 clusters which have internal user overlap of 15-20\%, and external overlap of 0-5\%.
The presence of these clusters implies the existence of hierarchical conversational contexts. 
Prior work in story detection, which the present work builds on, have also acknowledged the hierarchical nature of stories \cite{srijith2017sub}.
Here, we identify 2-layers of this hierarchy, which is of use when qualitatively analyzing the data within these contexts.
Future work that explicitly models the hierarchical nature of conversational contexts is of interest.

Next, we find that combined network analysis of multiple contexts severely corrupts measurements of important actors.
While the introduction of additional edges into a network can be expected to change network characteristics, some nodes drop from being a top-5 node in the contextualized network to being in the bottom quartile of central nodes in the combined network.
The severity of these changes implies that contextual mixing may be ``favoring" nodes which are situated between contexts, regardless of their standing \textit{within} context.
This is sensible, as they can be brokers of information between two loosely connected networks.
So, central nodes in the combined network \textit{are} important, and this approach allows us to be more specific about why; they are important because of their position between conversations.
Centrality measures accounting for node position within and between contexts are thus an interesting area of future work.
These may be developed in a similar manner to community-aware centrality measures, which account for node position within and between communities \cite{magelinski2021measuring,rajeh2021characterizing}.

Lastly, we observe that conversational dynamics are complex. 
Users do not transition between contexts in a simple sequential manner.
As a result, some contexts can be seen as more important than others, drawing the attention and engagement of users from a variety of areas.
The method of observing these important conversational contexts is useful for qualitative data analysis, however quantitative methods of determining context's importance and relating that to the importance of users is also of interest for future work.

\section{Limitations and Future Work}
There are a number of limitations to this work.
First, the data was collected using a keyword-based collection, resulting in only a sample of the true conversational network.
Twitter's new API provides functionality to build out the full conversational graph after the initial sample is collected\footnote{\url{https://blog.twitter.com/developer/en_us/topics/tools/2020/introducing_new_twitter_api}}.
This is a powerful data collection tool that will be leveraged in future work; obtaining the full conversational network will provide more edges for information to flow between.

Additionally, the initial feature representation of tweets are derived from a relatively simple language embedding scheme which does not include attached images or video.
The scheme was selected due to its scalability and its ability to embed tweets written in different languages within the same vector space. 
This approach embedded tweets from 11 languages, but 4\% of the reachable tweets were still not reached.
The lack of image or video representation is the more important limitation, particularly because many tweets with images or video do not have text.
While this is largely the case for replies and not original tweets, the full space is affected due to the transfer of information from reply to base tweet and vice-versa.
Even though a pre-trained model can be used to obtain image or video representations, the process of including this information in initial tweet feature representation is unclear.
Most tweets lack images, so feature concatenation will not be effective.
Combining the feature vectors is also not straightforward because the vector spaces are not aligned.
A process which gives a feature representation of both text and images is left for future work.

All tweets are treated equally in current methods, however, social signals such as the number of retweets or favorites a tweet gets could inform more sophisticated aggregation schemes for GNNs. 
This is left for future work.
Further, methods which incorporate URLs domain name could improve embeddings but are also left for future work. 

Another limitation of the current analysis is the lack of mention representation.
Mentions are a core feature of Twitter, allowing for users to directly tag other users in their Tweets.
Incorporating mention nodes into Deep Tweet Infomax should improve tweet representation, since  mentions are so commonly used to tag major players in a discussion.
This was not done in the present work because of the quality of the data available.
In the first version of Twitter's API, replying to a tweet adds a ``mention" of the user that is being replied to, and often adds ``mentions" to several other user higher in the  conversation tree.
These are not actual ``mentions," just artifacts of the already modeled tweet-tweet graph, so their inclusion could harm our results.
This problem is resolved in the new API.
Future work applying DTI to datasets collected with the new API will model mentions in a similar way to hashtags and URLs. 

The last major limitations of the approach are that the method still derives \textit{discrete} conversational contexts and that these are compared with noisy human annotations.
Specifically, the fact that our annotations were given by a single annotator poses a limitation.
Next, we have seen that interactions can be represented as occurring in a \textit{continuous} context space. However, all existing network approaches assume discrete context spaces.
Given the appropriate methods, the continuous context space could be used to measure things such as conversational drift and contextual persistence of links.
Thus, methods directly operating in continuous space are of interest in future work.

\section{Broader Perspectives and Ethics}
Contextualizing data allows for more accurate representation of user's importance within a discussion.
Social media analysis can have high stakes when it is used to determine the importance, or presence, of users within information operations.
While this work moves closer to properly attributing users to the conversations that they are actually active in, there is a question of interpretability.
The move towards deep graph neural networks makes interpretability challenging.
While the initial layer of our model is easy to interpret, this becomes more challenging as layers are stacked.
We have tried to validate that our model is appropriately representing the data by checking even the intermediate node representations of hashtags and URLs, but if this work is to be applied to qualitative work looking to attribute accounts a high-stakes setting, much more in-depth checks about how specific users fit into a conversation must be taken.

\section{Conclusion}
We provide a method of contextualizing conversational networks on Twitter.
This method represents tweets in a vector space using their text, hashtags, URLs, and the conversational network.
Vectorized tweets are then clustered into conversational contexts.
We apply our approach to a dataset of 4.5 million tweets and validate the results through inspection of nearest-neighbors in the embedded space, and by comparison with a label propagation procedure.

Conversational contexts have been shown to low overlap in user participation.
Thus, distinguishing between contexts allows for more accurate analysis about which users are participating in the same discussions.
Further, we see that it allows for more accurate analysis of which users are important within conversations, as contextualized networks are then demonstrated to have different central actors.
This points to an area of future work: quantification of user importance within and between conversational contexts.

Lastly, contextualized discussions are seen to have rich dynamics.
The Markov modeling of contextual dynamics provided in this work allow for the identification of important conversational contexts and give a strong qualitative understanding of the conversational flow within the dataset.
This opens another area of future work, where the interplay between user centrality and contextual relevance is studied.

While we analyzed Twitter data, the approach is readily extensible to other mediums. We hope that this work has demonstrated the importance of uncovering the context in which social connections are made online and will spark future work both in detecting and understanding the implications of such contexts.



\bibliography{references}

\end{document}